**Numerical stability of solitons waves through splices in quadratic optical media**

**Camila Fogaça de Oliveira**[a,1], **Paulo Laerte Natti**[b,2,*], **Eliandro Rodrigues Cirilo**[b,3], **Neyva Maria Lopes Romeiro**[b,4] **and Érica Regina Takano Natti**[c,5]

[a]Serviço Nacional de Aprendizagem Industrial de Londrina – SENAI Londrina

Rua Belém, 844, Londrina, Paraná, Brasil, 86026-440

[b]Universidade Estadual de Londrina, Departamento de Matemática (DMAT/UEL)

C.P. 6001, Londrina, Paraná, Brasil, 86051-990

[c] Pontifícia Universidade Católica do Paraná

Rua Jóquei Clube, 458, Londrina, Paraná, Brasil, 86067-000

[1] E-mail: camila.oliveira@pr.senai.br

[2] Corresponding author. E-mail: plnatti@uel.br. Web page: http://www.mat.uel.br/plnatti

Tel: 55-43-33714226. Fax: 55-43-33714216

[3] E-mail: ercirilo@uel.br

[4] E-mail: nromeiro@uel.br

[5] E-mail: erica.natti@pucpr.br

**ABSTRACT**. The propagation of soliton waves is simulated through splices in quadratic optical media, in which fluctuations of dielectric parameters occur. A new numerical scheme was developed to solve the complex system of partial differential equations (PDEs) that describes the problem. Our numerical approach to solve the complex problem was based on the mathematical theory of Taylor series of complex functions. In this context, we adapted the Finite Difference Method to approximate derivatives of complex functions and resolving the algebraic system resulting from the discretization, implicitly, by means of the relaxation Gauss-Seidel method. The mathematical modeling of local fluctuations of dielectric properties of optical media was performed by Gaussian functions. By simulating soliton wave propagation in optical fibers with Gaussian fluctuations in their dielectric properties, it was observed that the perturbed soliton numerical solution presented higher sensitivity to fluctuations in the dielectric parameter $\beta$, a measure of the intensity of nonlinearity in the fiber. In order to verify whether the fluctuations of $\beta$ parameter in the splices of the optical media generate unstable solitons, the propagation of a soliton wave, subject to this perturbation, was simulated for large time intervals. Considering various geometric configurations and intensities of the fluctuations of parameter β, it was found that the perturbed soliton wave stabilizes, i.e., the amplitude of the wave oscillations decreases for increasing values of propagation distance. It is concluded that the propagation of perturbed soliton wave presents numerical stability when subjected to local Gaussian fluctuations (perturbations) of the dielectric parameters of the optical media.

Keywords: perturbed soliton propagation; optical media splice; numerical stability; Gaussian fluctuation; complex finite difference method; Gauss-Seidel method.

## Introduction

In context of optical communication via solitons, the experiments performed in the late nineties generated encouraging results. In 1998, Thierry Georges and his team at France Telecom, combining optical solitons of different wavelengths, demonstrated data transmission of 1 terabit per second (Guen et al., 1999), and in 2000, Algety Telecom, then located in Lannion, France, developed undersea telecommunication equipment for the transmission of optical solitons. However, these promising results were not translated into actual commercial soliton system deployments, in either terrestrial or submarine systems, chiefly due to the Gordon–Haus (GH) jitter effect (Okamawari, Maruta & Kodama, 1998). GH jitter requires a sophisticated and expensive compensatory solution that ultimately makes the Dense Wavelength Division Multiplexing (DWDM) soliton transmission unattractive. Consequently, in the last decade the long-haul soliton transmission has remained as a subject of laboratory research. On the other hand, several solutions have been proposed to minimize the jitter effect, such as Raman fiber amplifiers (El-Halawany, 2011), tapering dispersion fiber spans (Gumenyuk, Thür, Kivistö & Okhotnikov, 2010), sliding frequency guiding filters (Ferreira, Facao, Latas & Sousa, 2005), in-line synchronous modulation (Liu et al., 2011), among others. A review of the basic physics of the dynamics of solitons and futuristic applications of solitons in optical communication can be found in (Hasegawa, 2000).

In recent years, an increase in the number of theoretical and experimental works on soliton communications, that aim to overcome the many well-known problems and improve the methods already proposed, was published. Such studies approach themes related to the new soliton generation processes [Malomed, Mihalache, Wise & Torner 2005; Kurokawa, Tajima & Nakajima, 2007), soliton propagation processes (Latas & Ferreira, 2007; Tsaraf &

Malomed, 2009) and soliton stabilization processes (Biswas, 2002; Driben & Malomed, 2007; Chen & Malomed , 2009) in optical media.

This work is about propagation and stability of solitons in optical media. The propagation of these waves in optical media is affected by several disturbing processes. Usually, the most important ones are group velocity dispersion (chromatic dispersion) and optical Kerr effect (intensity dependence of the refractive index). Under certain circumstances, however, the effects of Kerr nonlinearity and dispersion can just cancel each other, so that the temporal and spectral shapes of the pulses is preserved even over long propagation distances. Taking only these disturbances into account, the pulse propagation is a soliton described by a system of coupled nonlinear Schrödinger differential equations (Hasegawa & Tappert, 1973; Menyuk, Schiek & Torner, 1994). The most remarkable fact about soliton waves is, actually, not the possibility of balance of dispersion and nonlinearity, but rather that soliton solutions of nonlinear wave equation are very stable: even for substantial deviations of the initial pulse from the exact soliton solution, the pulse automatically finds the correct soliton shape.

On the other hand, to describe real-world fiber-optic systems, it is more realistic to include further disturbing effects such as influence of fusion splice (Yin, Yan & Gong, 2011), Rayleigh scattering (Böhm & Mitschke, 2007), high-order dispersion and high-order nonlinearities (Agrawal, 2001), soliton self-steepening, Raman effect and self-frequency shift (Latas & Ferreira, 2007), polarization-mode dispersion (Driben & Malomed, 2007), nonlinear phase noise (Lau & Kahn, 2007), among others (Kohl, Biswas, Milovic & Zerrad, 2008).

It should be observed that the perturbed coupled nonlinear Schrödinger differential equations systems, which describe wave propagation in real optical media, do not present analytical solution. In the literature there are several numerical approaches whose objective is to describe the propagation of perturbed solitons in dielectric environments, most of them using the finite difference method (Ismail & Alamri, 2004; Wang, 2005; Chen & Malomed, 2009) or the finite element method (Dag, 1999; Ismail, 2008). On the other hand, to solve numerically the resulting system of equations, the authors use various methods like Newton's method (Ismail, 2008), Crank-Nicolson method (Chen & Malomed, 2009), Runge-Kutta Method (Reich, 2000), among others. A review of the several numerical procedures applied to describe the propagation of solitons in optical fibers is found in (Dehghan & Taleei, 2010).

In a previous work (Cirilo, Natti, Romeiro, Takano Natti & Oliveira, 2010), we described the propagation of soliton waves in ideal quadratic optical media through a procedure based on the Complex Finite Difference method and relaxation Gauss-Seidel method. By comparing the obtained numerical results with the known analytical results, the validation of the developed numerical procedure has been verified.

In this work, aiming to study the propagation and the stability of perturbed soliton waves, through splices in $\chi^{(2)}:\chi^{(2)}$ optical media, the general numerical procedure developed in (Cirilo et al., 2010) is used. In Section 2 we present the soliton analytical solutions of the coupled nonlinear differential equations system that described the propagation of soliton waves in materials with a cascaded $\chi^{(2)}:\chi^{(2)}$ nonlinearity. In Section 3, the numerical procedure to study the propagation of the perturbed solitons in optical media is described. In Section 4 the propagation of perturbed soliton waves is simulated through splices in optical media, where fluctuations of the dielectric parameters occur. The mathematical modeling of these local fluctuations of the dielectric properties was performed by Gaussian functions with various geometric configurations and intensities. In Section 5 the main results of this work are presented.

## Material and Methods

### Solitons in ideal dielectric planar waveguide

This section studies the coupled non-linear complex partial differential equations (PDE) system, obtained from Maxwell's equations, which describe the longitudinal propagation of two coupled electromagnetic waves (fundamental and second harmonic modes) in ideal $\chi^{(2)}$ dielectric planar waveguide. The detailed mathematical modeling of this PDE system can be found in (Menyuk, et al., 1994; Etrich, Lederer, Malomed, Peschel & Peschel, 2000; Buryak, Trapani, Skryabin & Trillo, 2002). This complex PDE system is given by

$$i\frac{\partial a_1}{\partial \xi} - \frac{r}{2}\frac{\partial^2 a_1}{\partial s^2} + a_1^{\,*}\, a_2 \exp(-i\beta\xi) = 0$$

$$i\frac{\partial a_2}{\partial \xi} - i\delta\frac{\partial a_2}{\partial s} - \frac{\alpha}{2}\frac{\partial^2 a_2}{\partial s^2} + a_1^{\,2} \exp(i\beta\xi) = 0 \ , \qquad (1)$$

where $i=\sqrt{-1}$ is the imaginary unit, $a_1(\xi,s)$ and $a_2(\xi,s)$ are complex variables that represent the normalized amplitudes of the electrical fields of the fundamental and second harmonic waves, respectively, with $a_1^*(\xi,s)$ and $a_2^*(\xi,s)$ as their complex conjugates.

The real parameters $\alpha$, $\beta$, $\delta$ and $r$, in (1), are related with the dielectric properties of the optical media (Menyuk, et al., 1994; Galleas, Ymai, Natti & Takano Natti, 2003). The $\beta$ quantity is a measure of the intensity of nonlinearity in the optical fiber, or a measure of the generation rate of the second harmonic. The $\alpha$ quantity measures the relative dispersion of the group velocity dispersion (GVD) of fundamental and second harmonic waves in the optical fiber. For values $|\alpha| > 1$, the second harmonic wave has higher dispersion than the fundamental wave and for values $|\alpha| < 1$, it is the fundamental wave that has higher dispersion. The $r$ quantity is the signal of the fundamental GVD wave. When $r = +1$, the fundamental wave is in normal dispersion regime, but if $r = -1$, the fundamental wave is in anomalous dispersion regime. Finally, parameter $\delta$ measures the difference of group velocities of fundamental and second harmonic waves, so it accounts for the presence of Poynting vector walk-off that occurs in birefringent media. It should be noticed that is possible to choose the characteristics (velocity, width, amplitude, etc.) of the wave to be propagated in the optical media, selecting or proposing materials with the appropriate $\alpha$, $\beta$, $\delta$ and $r$ dielectric properties.

The PDE system (1) presents soliton solutions. It is observed that the original idea of cascading due to second harmonic generation was proposed by Ostrovskii (1967), and Karamzin and Sukhorukov (1975). In this work we use cascaded $\chi^{(2)} : \chi^{(2)}$ nonlinearity soliton solutions as given by ((Menyuk, et al., 1994; Galleas, et al., 2003)

$$a_1(\xi, s) = \pm \frac{3}{2(\alpha - 2r)} \sqrt{\alpha\, r}\; \left(\frac{\delta^2}{2\alpha - r} + \beta\right) \times \operatorname{sech}^2 \left[ \pm \sqrt{\frac{1}{2(2r - \alpha)} \left(\frac{\delta^2}{2\alpha - r} + \beta\right)}\; \left(s - \frac{r\delta}{2\alpha - r}\xi\right) \right]$$

$$\times \exp\left\{ i \left[ \left( \frac{r\delta^2 (4r \quad 5\alpha)}{2(2\alpha - r)^2 (2r - \alpha)} - \frac{r\beta}{2r - \alpha} \right) \xi - \left( \frac{\delta}{2\alpha - r} \right) s \right] \right\} \tag{2}$$

$$a_2(\xi, s) = \frac{3r}{2(\alpha - 2r)} \left(\frac{\delta^2}{2\alpha - r} + \beta\right) \times \operatorname{sech}^2 \left[ \pm \sqrt{\frac{1}{2(2r - \alpha)} \left(\frac{\delta^2}{2\alpha - r} + \beta\right)}\; \left(s - \frac{r\delta}{2\alpha - r}\xi\right) \right]$$

$$\times \exp\left\{ 2i \left[ \left( \frac{r\delta^2 (4r \quad 5\alpha)}{2(2\alpha - r)^2 (2r - \alpha)} - \frac{r\beta}{2r - \alpha} + \frac{\beta}{2} \right) \xi - \left( \frac{\delta}{2\alpha - r} \right) s \right] \right\}, \tag{3}$$

where the term $v = r\,\delta / (2\alpha - r)$ is the velocity of these solitons, so the independent variable $s$ has spatial character, whereas the independent variable $\xi$ has temporal character.

In (Cirilo, et al., 2010) a numerical procedure was developed based on the Complex Finite Difference method and relaxation Gauss-Seidel method to solve the propagation of soliton waves in optical media described by the PDE system (1). By comparing the obtained numerical results with the known analytical solutions (2-3), the validation of the developed numerical procedure was verified. This numerical development is presented in the next section.

## Numerical model for the propagation of solitons in optical fibers

The system (1), which describes the propagation of solitons, is complex. Our numerical approach to solve the complex problem was based on the mathematical theory of Taylor series of complex functions. In this context, we adapted the Finite Difference Method to approximate derivatives of complex functions. We also decided to make approximations in order to obtain an Implicit Method, because the resulting linear system (in complex variables) became well-conditioned. Finally, due to ease of implementation, we chose to solve the resulting linear system by the Relaxation Gauss-Seidel Method (Smith, 2003; Sperandio, Mendes & Monken, 2003), which accelerates the convergence. We show that our procedure converges.

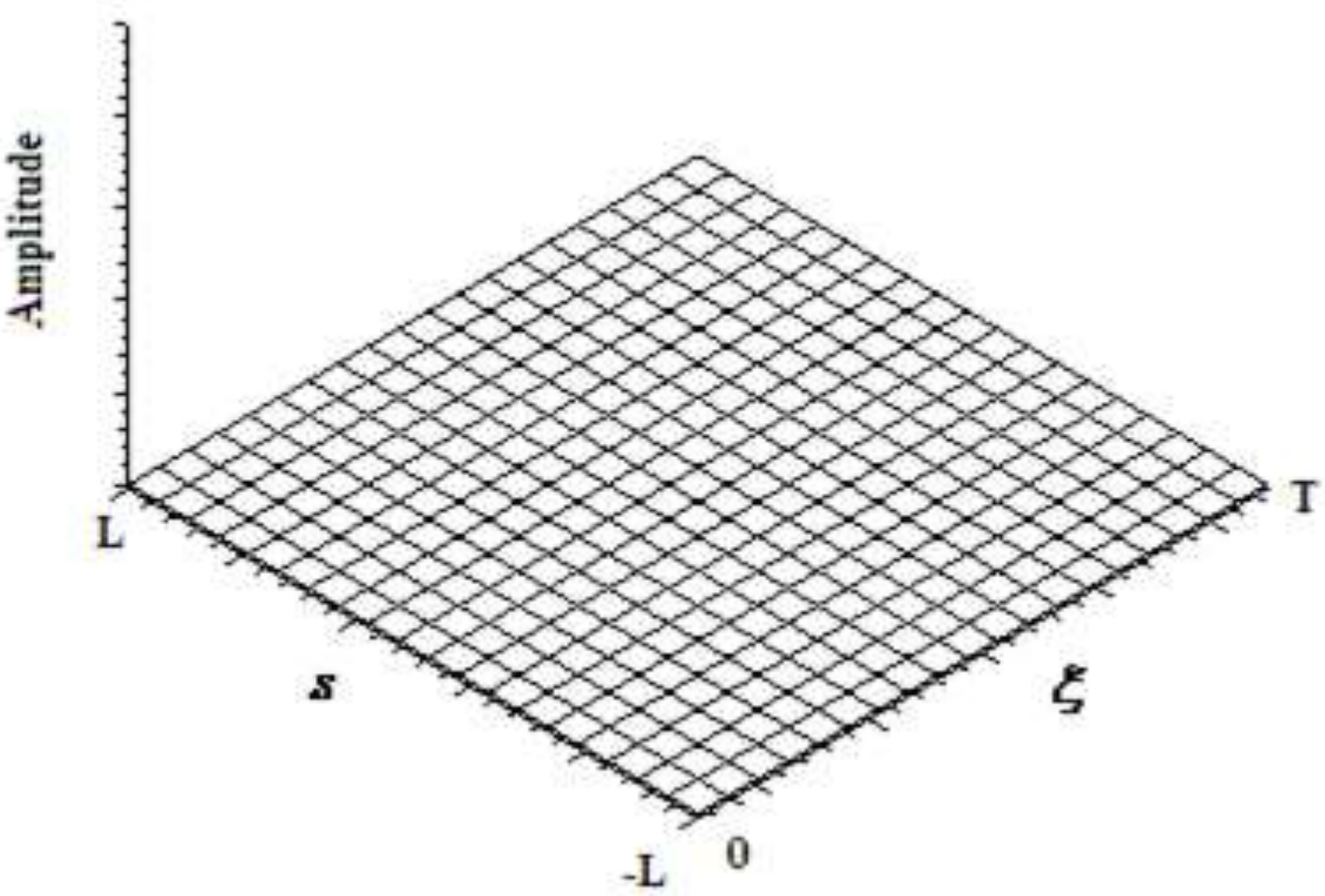

**Fig. 1.** Computational domain of the propagation of soliton waves.

The system (1) is numerically resolved in domain $\xi \times s = [0,T] \times [-L,L]$, where $T, L \in \Re$. By discretizing the variables $a_1(\xi,s) \equiv a_{1_{k+1,j}}$ and $a_2(\xi,s) \equiv a_{2_{k+1,j}}$, for $k = 0,1,...,k_{\max}$ and $j = 1,2,...,nj$, where $k_{\max}$ is denominated the last advance in $\xi$ and $nj$ the maximum number of points in $s$, the propagation domain of the soliton waves is defined by a discretized computational grid of $k_{\max} \times nj$ points, as represented in figure 1.

Thus, by means of the Complex Finite Difference Method developed, approaching the temporal derivates by progressive differences, and the spatial derivates by central differences (Smith, 2003), the following linear systems are generated from the differential equations (1), namely,

$$a_{1_{k+1,j}} = \frac{1}{{}^1A_p} [\ {}^1A_W a_{1_{k+1,j-1}} + {}^1A_E a_{1_{k+1,j+1}} + {}^1A_{po} a_{1_{k,j}} - a^*_{1_{k,j}} a_{2_{k,j}} \exp(-i\beta t_k)]$$

$$\tag{4}$$

$$a_{2_{k+1,j}} = \frac{1}{{}^2A_p} [\ {}^2A_W a_{2_{k+1,j-1}} + {}^2A_E a_{2_{k+1,j+1}} + {}^2A_{po} a_{2_{k,j}} - (a_{1_{k+1,j}})^2 \exp(i\beta t_k)]$$

where

$${}^1A_p = \frac{i}{\Delta\xi} + \frac{r}{(\Delta s)^2} \qquad {}^1A_E = {}^1A_W \qquad {}^1A_W = \frac{r}{2(\Delta s)^2} \qquad {}^1A_{po} = \frac{i}{\Delta\xi}$$

$${}^2A_p = \frac{i}{\Delta\xi} + \frac{\alpha}{(\Delta s)^2} \qquad {}^2A_E = \frac{i\delta}{2\Delta s} + \frac{\alpha}{2(\Delta s)^2} \qquad {}^2A_W = -\frac{i\delta}{2\Delta s} + \frac{\alpha}{2(\Delta s)^2}2 \qquad {}^2A_{p_0} = \frac{i}{\Delta\xi}$$

and in these approaches $k+1$ is the current time, $k$ is the previous time, $\Delta s$ is the spatial quantity discretized and $\Delta\xi$ is the temporal quantity discretized, so that $t_k = k\,\Delta\xi$.

In this work, the linear system (4) is resolved by means of the relaxation Gauss-Seidel method (Smith, 2003; Sperandio, Mendes & Monken, 2003; Cirilo, Natti, Romeiro & Takano Natti, 2008). Consider this linear system for $a_{1_{k+1,j}}$, given explicitly by

$$a_{1_{k+1,2}} = \frac{1}{{}^1A_p} [\ {}^1A_W a_{1_{k+1,1}} + {}^1A_E a_{1_{k+1,3}} + {}^1A_{po} a_{1_{k,2}} - a^*_{1_{k,2}} a_{2_{k,2}} \exp(-i\beta t_k)]$$

$$a_{1_{k+1,3}} = \frac{1}{{}^1A_p} [\ {}^1A_W a_{1_{k+1,2}} + {}^1A_E a_{1_{k+1,4}} + {}^1A_{po} a_{1_{k,3}} - a^*_{1_{k,3}} a_{2_{k,3}} \exp(-i\beta t_k)]$$

$$a_{1_{k+1,4}} = \frac{1}{{}^1A_p} [\ {}^1A_W a_{1_{k+1,3}} + {}^1A_E a_{1_{k+1,5}} + {}^1A_{po} a_{1_{k,4}} - a^*_{1_{k,4}} a_{2_{k,4}} \exp(-i\beta t_k)]$$

...

$$a_{1_{k+1,nj-1}} = \frac{1}{{}^1A_p} [\ {}^1A_W a_{1_{k+1,nj-2}} + {}^1A_E a_{1_{k+1,nj}} + {}^1A_{po} a_{1_{k,nj-1}} - a_1^*(k,nj-1) a_2(k,nj-1) \exp(-i\beta t_k)].$$

It can be written in compact form as

$$a_{1_{k+1,j}} = \frac{{}^{1}B_{k,j} + {}^{1}A_W a_{1_{k+1,j-1}} + {}^{1}A_E a_{1_{k+1,j+1}}}{{}^{1}A_p} \quad ,$$

where ${}^{1}B_{k,j} = {}^{1}A_{po} a_{1_{k,j}} - a^{*}_{1_{k,j}} \; a_{2_{k,j}} \exp\left(-i\beta t_k\right)$ with $j = 2,3,\ldots,nj\text{-}1$.

From the initial condition $a_{1_{0,j}}$, given by soliton solution (2), and imposing the contour conditions $a_{1_{k+1,1}} = 0$ and $a_{1_{k+1,nj}} = 0$, for $L$ sufficiently large, $a_{1_{k+1,j}}{}^{(l)}$ is iteratively calculated by means of the equations

$$a_{1_{k+1,j}}{}^{(l)} = \frac{{}^{1}B_{k,j}{}^{(l)} + {}^{1}A_W a_{1_{k+1,j-1}}{}^{(l)} + {}^{1}A_E a_{1_{k+1,j+1}}{}^{(l-1)}}{{}^{1}A_p} \quad , \tag{5}$$

where $l = 1,2,3\ldots$ is the iterative level, ${}^{1}B_{k,j}{}^{(l)} = {}^{1}A_{po} a_{1_{k,j}}{}^{(l)} - a^{*}_{1_{k,j}}{}^{(l)} \; a_{2_{k,j}}{}^{(l)} \exp\left(-i\beta t_k\right)$, until the stop criterion is fulfilled, namely,

$$\max_{2 \le j \le nj-1} \left| a_{1_{k+1,j}}{}^{(l)} - a_{1_{k+1,j}}{}^{(l-1)} \right| < 10^{-6} \; . \tag{6}$$

This method consists in determining $a_{1_{k+1,j}}{}^{(l)}$ by using the already known components of $a_{1_{k+1,j+1}}{}^{(l-1)}$ and $a_{1_{k+1,j-1}}{}^{(l)}$, with the advantage of not requiring the simultaneous storage of the two vectors $a_{1_{k+1,j+1}}{}^{(l-1)}$ and $a_{1_{k+1,j-1}}{}^{(l)}$ at each step. Likewise, $a_{2_{k+1,j}}{}^{(l)}$ is resolved.

It should be noticed that in equations (5)-(6) the value $w = 1.0$ was used for the parameter of relaxation (Cirilo et al., 2008). Such value corresponds to the optimal relaxation parameter in relation to the variations of the dielectric parameters $\alpha$, $\beta$ and $\delta$ of system (1). Figure 2 presents the flowchart of the numerical code developed for PDEs system (1).

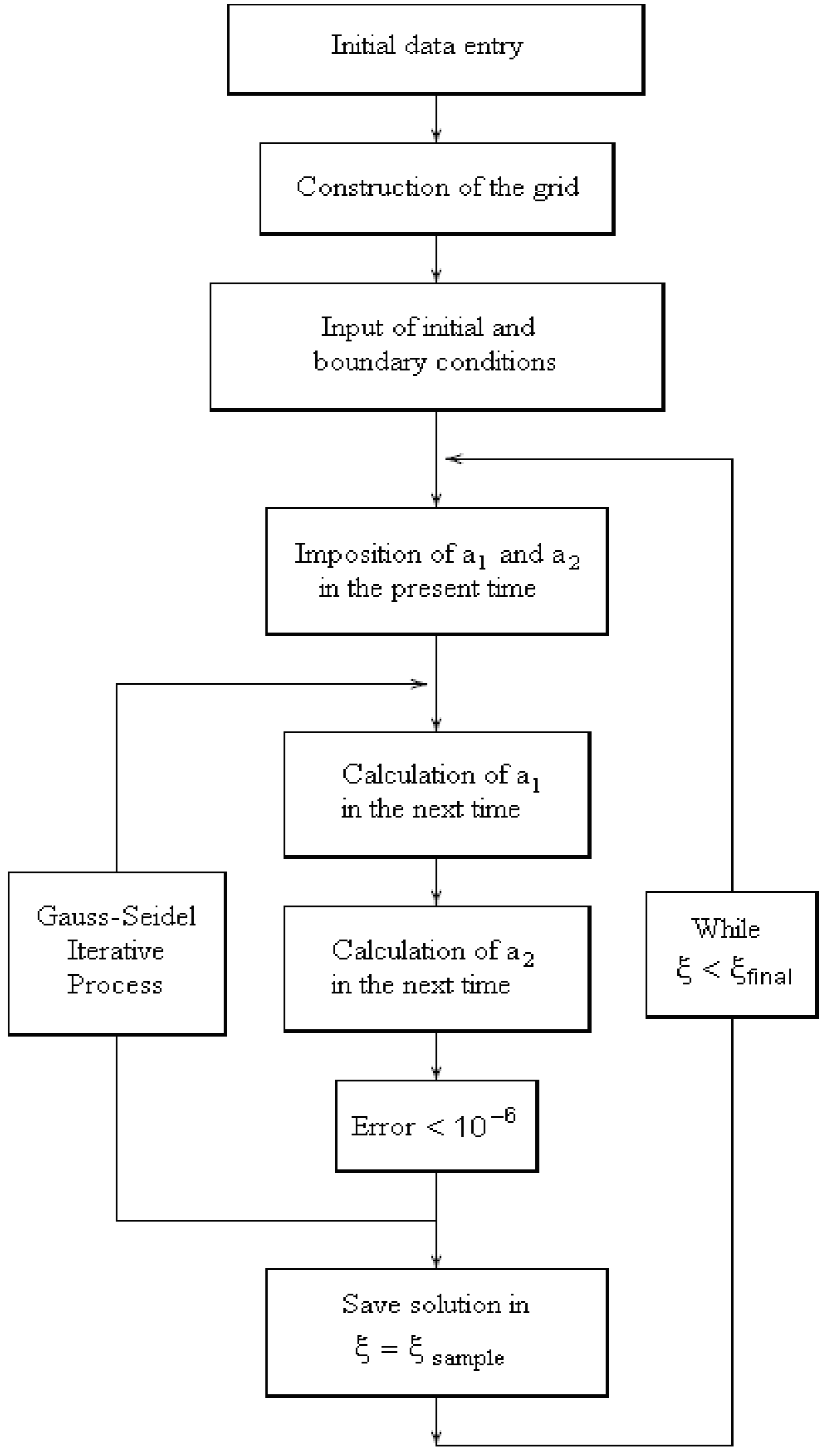

**Fig. 2.** Flowchart of the numerical code developed to obtain the numerical soliton solutions.

## Results and discussion

### Stability of perturbed soliton waves

In our previous work (Cirilo et al., 2010) the dielectric parameters $\alpha$, $\beta$ and $\delta$ in (1) were considered constant, so that PDE system (1) presented soliton solutions. On the other hand, in real optical media, the dielectric parameters $\alpha$, $\beta$ and $\delta$ are not constant. This section will study how localized fluctuations of the dielectric parameters occurring in the splices of optical media affect the propagation of soliton waves.

In the context of optical fibers, the fusion splicing process of optical fibers is usually realized by means of an electric arc, but it can be realized by laser, gas flame, or tungsten filament through which current is passed. The fusion splicing apparatus consists of two fixtures on which the fibers are mounted. The fibers in the apparatus are aligned and then fused. In fusion splicing, the splice loss is a direct function of the angles of alignment and quality of the two fiber-end-faces. A splice loss under 0.1 dB is typical. Alternatives to fusion splicing includes using optical fiber connectors or mechanical splices, both of which have higher insertion losses, lower reliability and higher return losses than fusion splicing.

In the fusion splicing process, the local dielectric properties of the optical media are modified along the spatial coordinate *s*, ranging from the default value to a maximum variation, in the region where the optical properties are more affected by the fusion splicing process, to then decrease again to the default value. In order to model these perturbations of optical properties in the fusion splicing regions, we will use a Gaussian function. To analyse the stability of the perturbed soliton waves as a function of the fluctuations of the dielectric parameters in the optical media is our aim. In this case, only numerical solutions are possible, since the propagated waves are not solitons given by (2-3) anymore.

In reference (Cirilo et al., 2010) we validate our numerical procedure by comparison between the numerical and analytical solution by getting smaller errors than $10^{-6}$. See equations (4)-(5) of this work.

### A localized perturbation in the optical fiber

Initially, the discretization of the computational grid will be described. For spatial variable $s$, the interval $-50 < s < 90$ was established, with discretization $\Delta s = 1.0 \times 10^{-1}$, whereas for temporal variable $\xi$, the interval was $0 < \xi < 50$, with discretization $\Delta\xi = 1.0 \times 10^{-3}$. The geometry of the computational grid was adjusted so that the physics of the propagation of the perturbed soliton wave is within the considered computational domain.

In the sequence, the mathematical modeling of the fluctuation of the dielectric parameters in the splices of the optical media will be considered, as well as such localized fluctuation affect the stability of the propagation of waves. In the modeling of the optical properties along the dielectric media used in this work, it is supposed that, in average, the dielectric parameters take values $r = -1.0$, $\alpha = -1/4$, $\beta = -1/2$ and $\delta = -1/4$. It is also supposed that in the areas surrounding the splices of two optical media, the optical properties $\alpha$, $\beta$ and $\delta$ are altered according to a Gaussian function, as explained above. In the following simulations, Gaussian fluctuations of 5% in the values of the dielectric parameters are considered.

**Parameter** $\alpha$: Figure 3 shows the simulations of perturbations in $a_1(\xi, s)$ and $a_2(\xi, s)$ due to the fluctuations in parameter $\alpha$, around $\alpha = -1/4$, with maximum amplitude corresponding to 5% of the average value of the dielectric parameter. By establishing values $r = -1.0$, $\beta = -1/2$, $\delta = -1/4$ and varying $\alpha$ by means of the Gaussian function $\alpha = (-1/4)\,[1 + 0.05 \times \exp(-(\xi - 10)^2)]$, there is the occurrence of small variations in the amplitudes of $a_1(\xi, s)$ and $a_2(\xi, s)$, not easily visualized in figure 3. From the Gaussian function, it is observed that the fluctuations in $\alpha$ occur around $\xi = 10$.

**Parameter** $\delta$: Figure 4 shows the simulations of perturbations in $a_1(\xi, s)$ and $a_2(\xi, s)$ due to the fluctuations in parameter $\delta$, around $\delta = -1/4$, with maximum amplitude corresponding to 5% of the average value of the dielectric parameter. By establishing values $r = -1.0$, $\alpha = -1/4$, $\beta = -1/2$ and varying $\delta$ by means of the Gaussian function $\delta = (-1/4)\,[1 + 0.05 \times \exp(-(\xi - 10)^2)]$, there is the occurrence of small variations in the amplitudes of $a_1(\xi, s)$ and $a_2(\xi, s)$, not easily visualized in figure 4. From the Gaussian function, it is observed that the fluctuations in $\delta$ occur around $\xi = 10$.

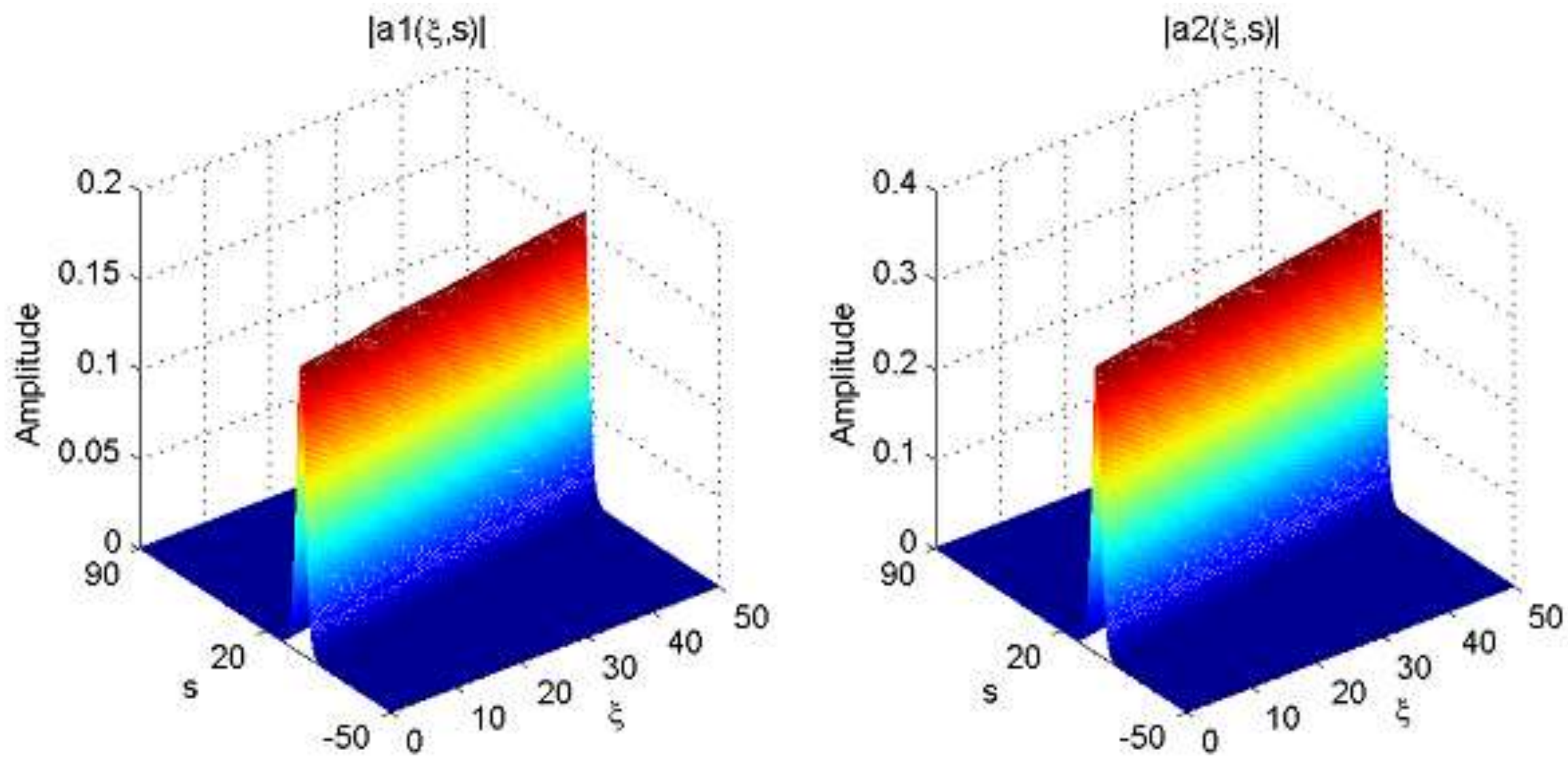

**Fig. 3.** Numerical solutions $a_1(\xi,s)$ and $a_2(\xi,s)$ when $\alpha = (\text{-}1/4)\,[1+0.05\times\exp(\text{-}(\xi\,\text{-}10)^2)]$ with $r = \text{-}1.0$, $\beta = \text{-}1/2$ and $\delta = \text{-}1/4$.

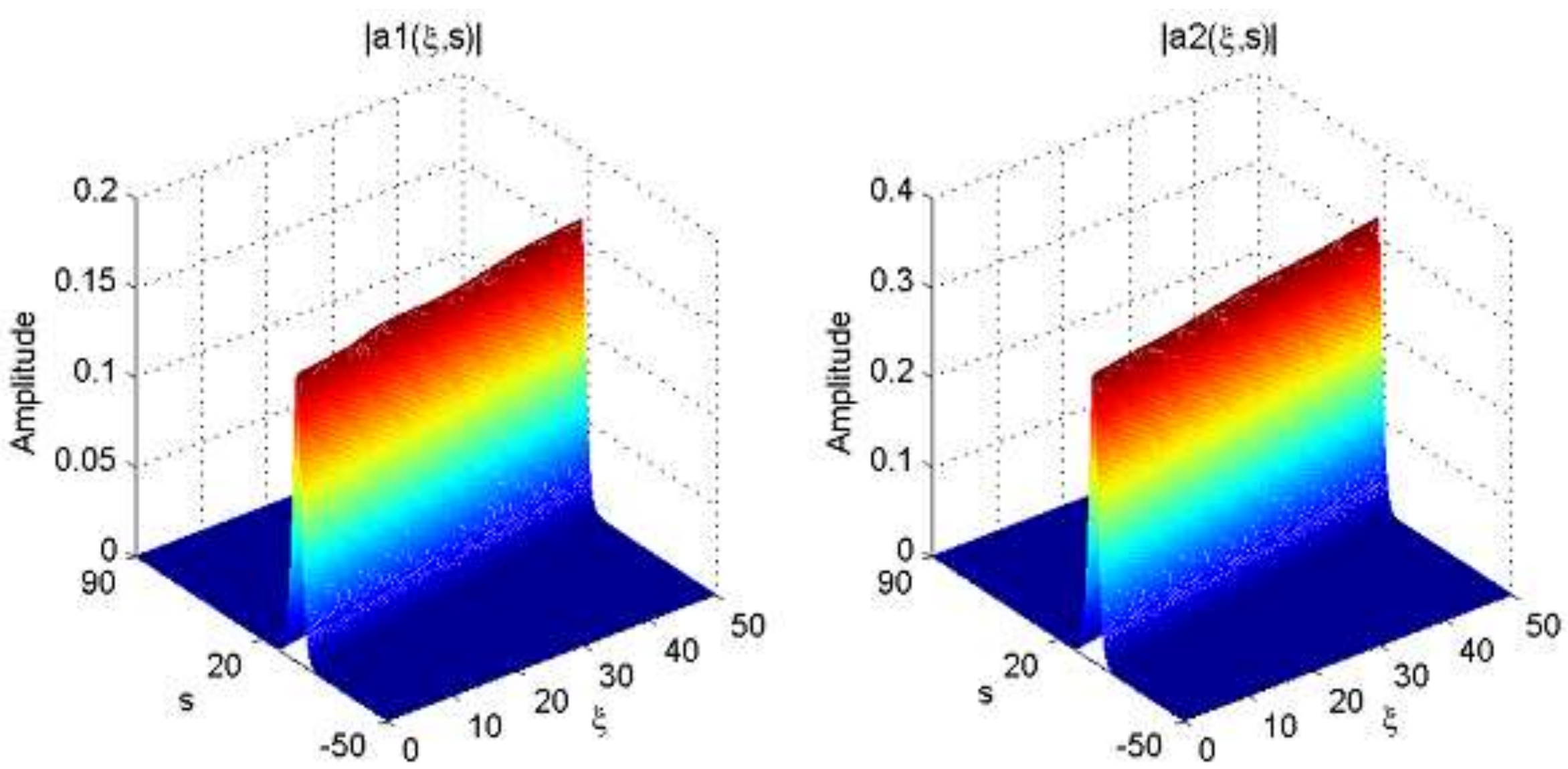

**Fig. 4.** Numerical solution $a_1(\xi,s)$ and $a_2(\xi,s)$ when $\delta = (\text{-}1/4)\,[1+0.05\times\exp(\text{-}(\xi\,\text{-}10)^2)]$ with $r = \text{-}1.0$, $\alpha = \text{-}1/4$ and $\beta = \text{-}1/2$.

**Parameter** $\beta$ : Figure 5 shows the simulations of perturbations in $a_1(\xi,s)$ and $a_2(\xi,s)$ due to the fluctuation in parameter $\beta$, around $\beta = \text{-}1/2$, with maximum amplitude corresponding to 5% of the average value of the dielectric parameter. By establishing values $r = \text{-}1.0$, $\alpha = \text{-}1/4$, $\delta = \text{-}1/4$ and varying $\beta$ by means of the Gaussian function $\beta = (\text{-}1/2)\ [1+0.05\exp(\text{-}(\xi\text{-}10)^2)]$, in figure 5 there is the occurrence of visible variations in the amplitudes of $a_1(\xi,s)$ and $a_2(\xi,s)$. From the Gaussian function, it is observed that the fluctuations in $\beta$ occur around $\xi = 10$.

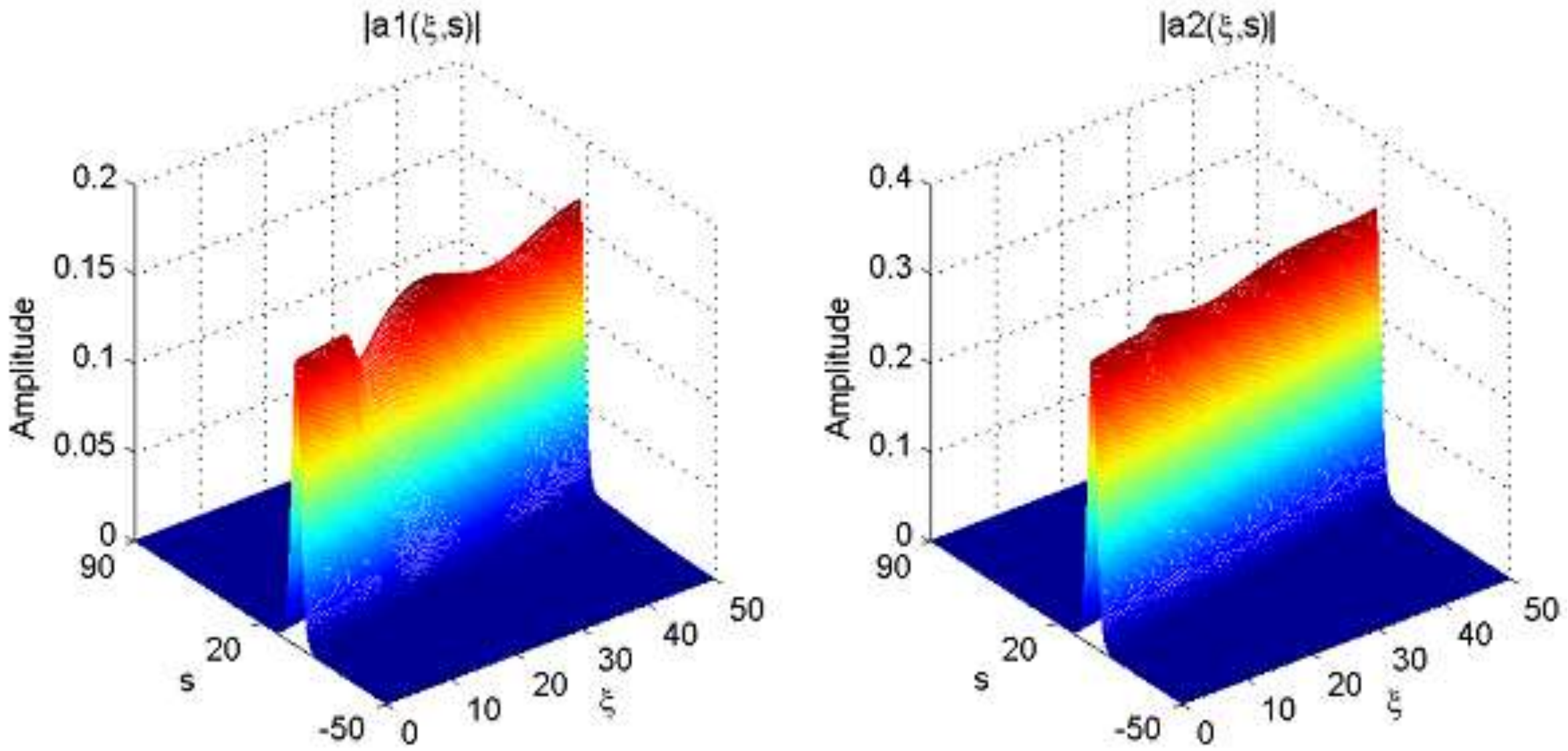

**Fig. 5.** Numerical solution $a_1(\xi,s)$ and $a_2(\xi,s)$ when $\beta = (\text{-}1/2)[1+0.05\times\exp(\text{-}(\xi\text{-}10)^2)]$ with $r = \text{-}1.0$, $\alpha = \text{-}1/4$ and $\delta = \text{-}1/4$.

When figures 3, 4 and 5 are compared, it is verified that the propagation of the soliton wave was more sensitive to the fluctuations of parameter $\beta$.

### Periodic perturbations

In this subsection, the computational grid was constructed by considering, for variable $s$, the interval $\text{-}50 < s < 90$, with discretization $\Delta s = 0.9\times 10^{-1}$, and for variable $\xi$, the interval $0 < \xi < 60$, with discretization $\Delta\xi = 1.0\times 10^{-3}$. Again, the geometry of the computational grid was adjusted so that the physics of the propagation of the perturbed soliton wave is completely within the considered computational domain.

It was also considered that the fluctuations of the dielectric parameters in the splices of the optical media are modeled by means of Gaussian functions. In the simulations conducted, periodic fluctuations of 1% were considered in the values of the dielectric parameters in $\xi = 5k$ with $k = 1,..,11$. By establishing, for the dielectric parameters, the average values $r = \text{-}1.0$, $\alpha = \text{-}1/4$, $\beta = \text{-}1/2$ and $\delta = \text{-}1/4$, and considering the described periodic Gaussian perturbations, it is verified again that $a_1(\xi,s)$ and $a_2(\xi,s)$ are more sensitive to the fluctuations of parameter $\beta$.

Figure 6 shows the simulations of perturbations in $a_1(\xi,s)$ and $a_2(\xi,s)$ due to the periodic fluctuations in parameter $\beta$, around $\beta = -1/2$, when $r = -1.0$, $\alpha = -1/4$, $\delta = -1/4$. In this case, significant variations are observed in the amplitudes of $a_1(\xi,s)$ and $a_2(\xi,s)$.

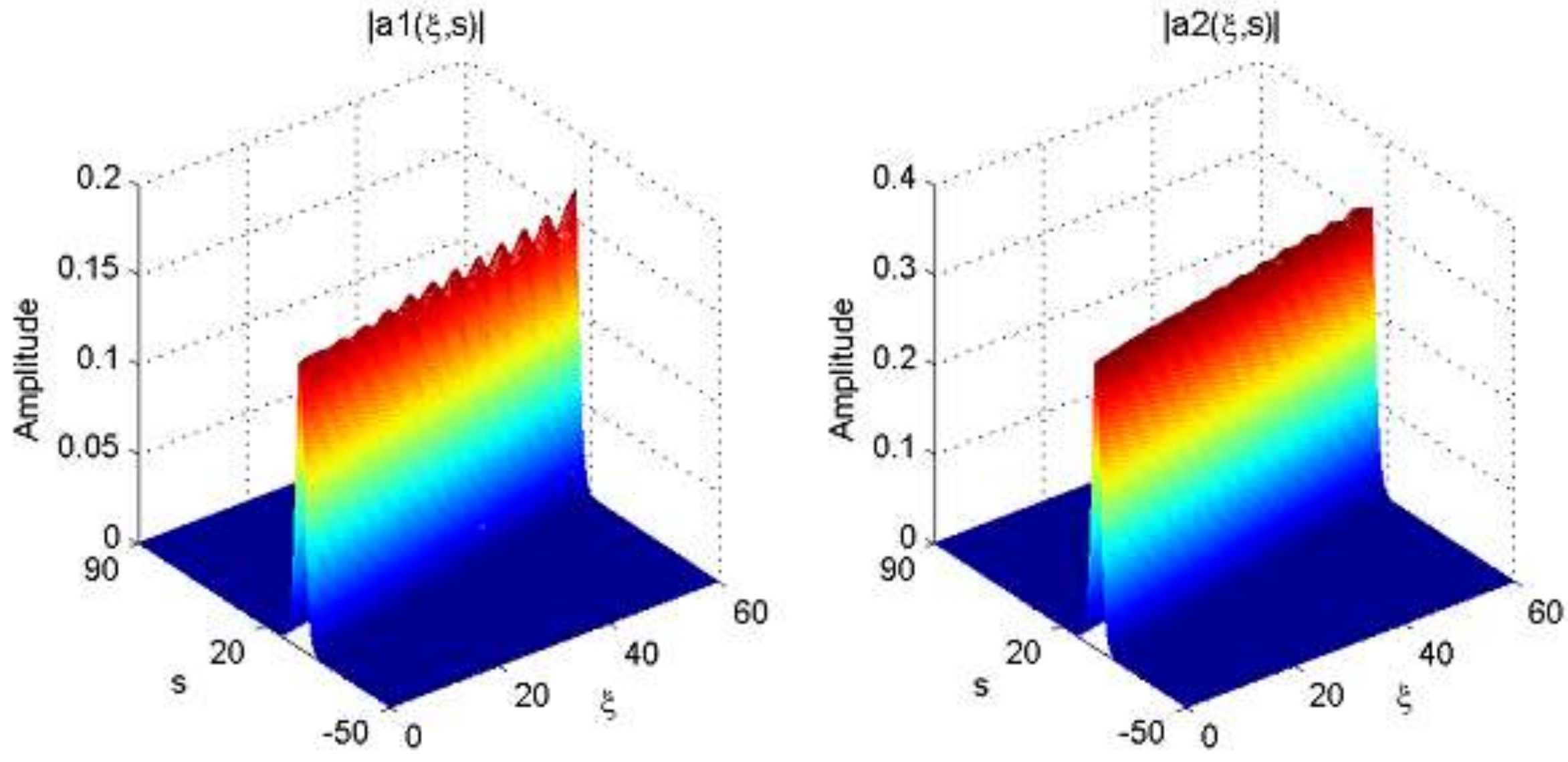

**Fig. 6.** Numerical solution $a_1(\xi,s)$ and $a_2(\xi,s)$ for 1% Gaussian perturbations in the $\beta$ value, in $\xi = 5k$, for $k = 1,..,11$, with $r = -1.0$, $\alpha = -1/4$, $\delta = -1/4$.

From the simulations in this subsection, it is observed that, in the context of optical communication with localized perturbative processes, the propagation of soliton waves is more affected by the fluctuations in parameter $\beta$. Therefore, the procedures for carrying out splices in optical media should be conceived so that, locally, the dielectric properties related to parameter $\beta$ are little affected.

### Parameter $\beta$ and the stability of perturbed solitons

In the preceding sections, by simulating the propagation of soliton waves through localized perturbations with various geometric configurations and intensities, it was verified that the soliton wave is more distorted when there are fluctuation in parameter $\beta$. With the objective of verifying whether the perturbed soliton waves achieve stability in function of the fluctuations in the dielectric parameter $\beta$, in the sequence, the evolution of such waves with higher values of temporal coordinate $\xi$ will be studied. Thus a computational grid was constructed, considering for variable $s$ the interval $-50<s<90$, with discretization $\Delta s = 5.6\times10^{-2}$, and for variable $\xi$, the interval $0<\xi<100$, with discretization $\Delta\xi = 1.0\times10^{-3}$. In all the following simulations, the values considered for the dielectric parameters were: $r = -1.0$, $\alpha = -1/4$, $\delta = -1/4$.

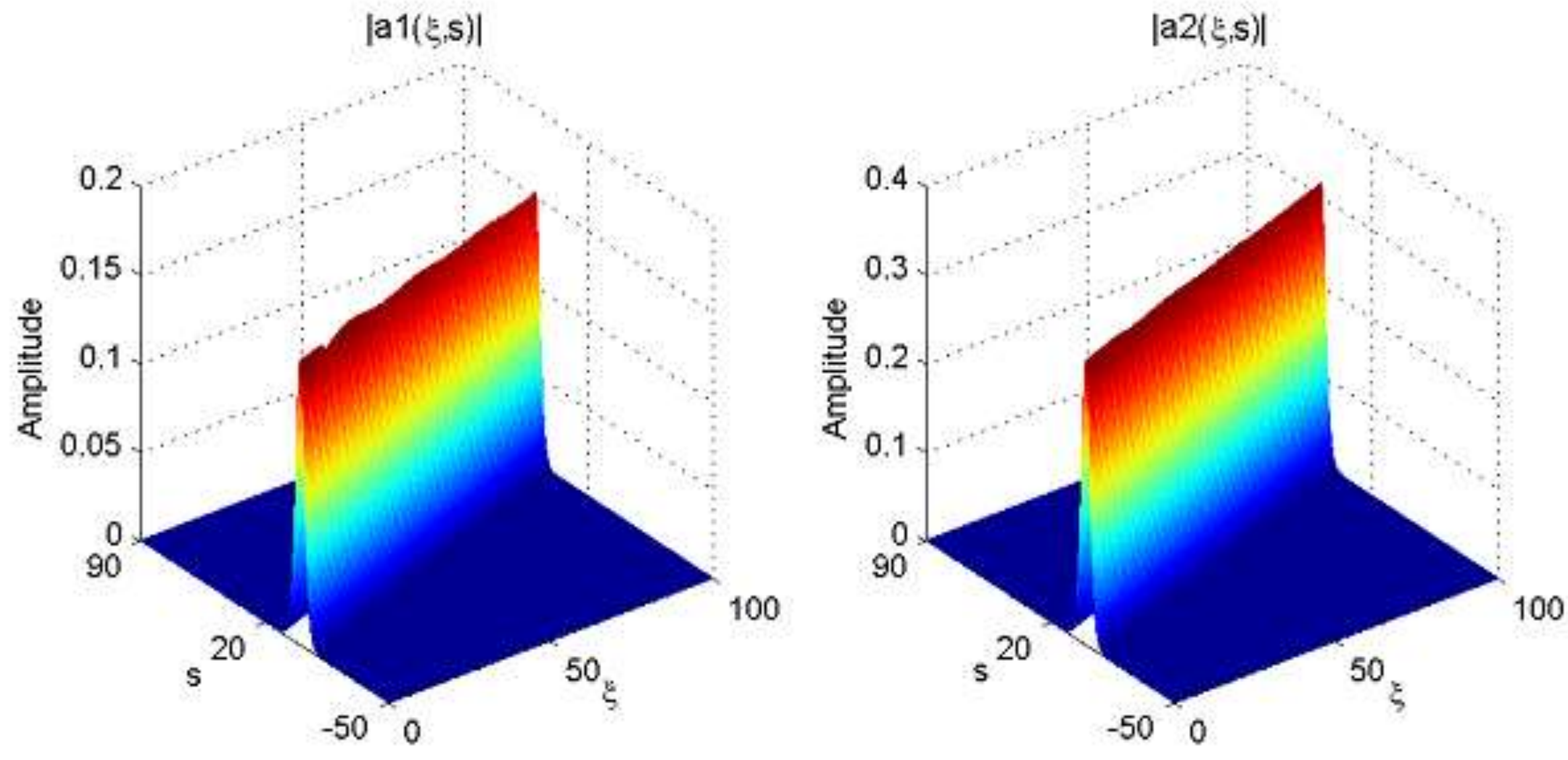

**Fig. 7.** Numerical solution $a_1(\xi,s)$ and $a_2(\xi,s)$ when $\beta = (-1/2)[1+0.01\times\exp(-(\xi-10)^2)]$ with $r = -1.0$, $\alpha = -1/4$ and $\delta = -1/4$, for large values of temporal coordinate $\xi$.

In figure 7, a Gaussian-type perturbation of 1% is considered in parameter $\beta$, around $\xi = 10$. It is verified that $a_1(\xi,s)$ and $a_2(\xi,s)$ evolve towards a stationary situation for higher values of $\xi$.

In figure 8, a Gaussian-type perturbation of 5% is considered in parameter $\beta$, around $\xi = 10$. It is observed again that the perturbed soliton wave evolves towards a stationary situation with damped fluctuations for increasing values of $\xi$.

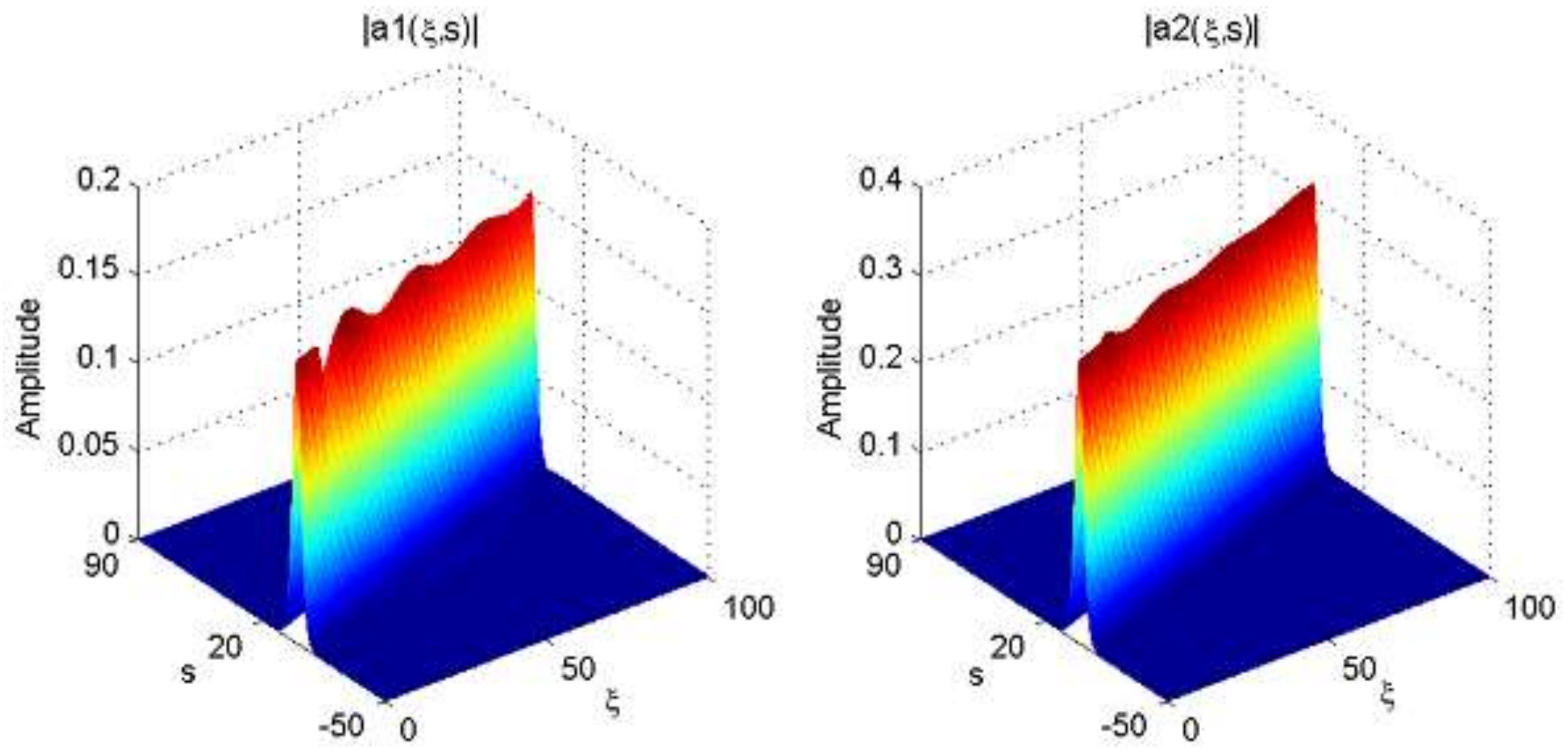

**Fig. 8.** Numerical solution $a_1(\xi,s)$ and $a_2(\xi,s)$ when $\beta = (-1/2)\,[1 + 0.05 \times \exp(-(\xi - 10)^2)]$ with $r = -1.0$, $\alpha = -1/4$ and $\delta = -1/4$, for large values of temporal coordinate $\xi$.

## Conclusions

In this work, the stability of the propagation of soliton waves through $\chi^{(2)}$ optical media splices was studied. It is observed that the dielectric properties are locally altered in splicing process of optical media. In order to simulate the propagation of solitons through such splices, considered as perturbations, the fluctuations of the dielectric parameters were locally modeled by means of Gaussian functions. By considering local and periodic configurations for the optical media splices, it was verified that the perturbed soliton wave presents higher sensitivity to parameter $\beta$, a measure of the phase mismatch; in other words, fluctuations in the dielectric parameter $\beta$ generate higher amplitude perturbations (oscillations) in the soliton waves. With the objective of verifying whether such perturbations generate unstable solitons, the evolution of the soliton wave was studied for several configurations of perturbations along of optical media, for higher values of temporal coordinate $\xi$. It was verified that, after the end of the perturbations, the perturbed soliton wave achieves stability, i.e., the amplitude of the oscillations decrease for increasing values of temporal coordinate $\xi$. It is therefore concluded that soliton waves, subject to Gaussian perturbations in the dielectric parameters of the optical media, present numerical stability.

As an experimental application of this work we propose that the procedures and experimental methods utilized in the process of optical media fusion are designed so that, locally, the $\beta$ dielectric properties are less altered during the splice process.

## Acknowledgements

We express our appreciation to the late Dr. Valdemir Garcia Ferrreira, whose contribution to this work was of great significance.